\documentclass[11pt,a4paper]{article}
\usepackage{graphicx}
\usepackage{amsmath}
\usepackage{caption}
\usepackage{subcaption}
\usepackage{framed,multirow}
\usepackage{xcolor}
\usepackage[margin=0.85in]{geometry} 
\begin{document}    
\title{Improving style transfer in dynamic contrast enhanced MRI using a spatio-temporal approach}

\author{Adam G. Tattersall, Keith A. Goatman, Lucy E. Kershaw,  \\
Scott I. K. Semple and Sonia Dahdouh} 
%
%\authorrunning{A. Tattersall et al.}
%\institute{University of Edinburgh, Edinburgh, UK \and
%Canon Medical Research Europe (CMRE), Edinburgh, UK}
%
%\titlerunning{Style transfer in dynamic contrast enhanced MRI}
\date{\vspace{-1.5ex}}
\maketitle              % typeset the header of the contribution
\begin{abstract}

Style transfer in DCE-MRI is a challenging task due to large variations in contrast enhancements across different tissues and time. Current unsupervised methods fail due to the wide variety of contrast enhancement and motion between the images in the series. We propose a new method that combines autoencoders to disentangle content and style with convolutional LSTMs to model predicted latent spaces along time and adaptive convolutions to tackle the localised nature of contrast enhancement. To evaluate our method, we propose a new metric that takes into account the contrast enhancement. Qualitative and quantitative analyses show that the proposed method outperforms the state of the art on two different datasets.

%\keywords{Style Transfer \and Spatio-temporal Information \and Content/Style Disentanglement \and Dynamic Contrast Enhanced Magnetic Resonance Imaging (DCE-MRI)}
\end{abstract}

\section{Introduction}
% Deep learning approaches to image processing require large datasets (e.g. ImageNet \cite{ILSVRC15}) to train robust models, however medical imaging datasets are often smaller with limited annotations. Augmentation is often used to artificially enlarge the dataset to enable models to learn from more examples.

% Obtaining manually annotated ground truth for this type of data is time consuming and difficult. Due to these challenges, fully annotated DCE-MRI datasets are scarce, reducing the diversity of available data to train deep learning models. We previously found that to train a robust model with DCE-MRI data, strategically using images from the whole sequence with varying CE strategically, leads to better performing models \cite{spiepaper}. However, in practice this can be difficult for tasks such as segmentation when ground truth is limited. Although common augmentation techniques allow for an increase in the number of images, they do not help to synthesise new annotated examples with varying CE. There are methods which can generate new images, however, the generated images do not use any previously available ground truth. A method that can synthesise new images from previously annotated data would provide models with diverse examples.

DCE-MRI is a type of quantitative imaging that can be used to monitor microvascular perfusion \cite{Ingrisch2013}. Multiple T1 weighted images are taken rapidly over a few minutes whilst a contrast agent injection causes a rapid increase in signal intensity in tissues of interest. The resulting data contain motion and tissue-dependent contrast enhancement (CE) which makes obtaining manually annotated ground truth time consuming and difficult. Fully annotated datasets are therefore scarce, which reduces the diversity of available data to train deep learning models. We previously found that to train a robust model with DCE-MRI data, using images strategically from the whole sequence with varying contrast levels led to better performing models \cite{spiepaper}. However, in practice this can be difficult for tasks such as segmentation when ground truth is limited. Although common augmentation techniques allow for an increase in the number of images, they do not help to synthesise new annotated examples with varying CE. A method that can synthesise new images from previously annotated data would provide models with diverse examples.

In this work, we propose a new style transfer method to augment DCE-MRI with different contrast enhancement levels. Our method uses convolutional long short-term memory (LSTM) \cite{Chao2018} networks which leverage the temporal information from DCE-MRI to better predict structures of the content image and generate images with the noise characteristics of MRI whilst allowing addition or removal of contrast enhancement. We also propose a new metric, contrast weighted (CW) - structural similarity index measure (SSIM), that uses information on the localised aspect of tissue enhancement to evaluate either structural or style similarity between a generated and a content or style image, respectively.

\section{Related Work}
% Style transfer combines the structure of a content image with the style of another image to produce new images. 
Using neural networks for style transfer was first proposed by Gatys \textit{et al.} \cite{Gatys2016} which used a pre-trained VGG-Network \cite{Simonyan2015}. Instead of training, an iterative process of passing the content, style and output image through a pre-trained network was used. 
% The feature maps predicted by the pre-trained network were used to calculate a style and content loss. 
% The loss was then used to determine how much the pixels in the output image should be changed. 
Building on this success, improvements were made to speed up the iterative process by using a feed-forward network \cite{Ulyanov2016} as well as creating perceptual losses \cite{justin16} to give more visually pleasing results. However, these networks struggled to adapt to new styles that were introduced during inference as the network was usually tied to a small number of style images. In 2017, Huang \textit{et al.} \cite{Huang2017} proposed Adaptive Instance Normalisation (AdaIN) which aligns the mean and variance of the content features with those of the style features allowing for style transfer without the restriction of a pre-defined set of styles.

Successful style transfer methods have since been proposed such as learning mappings between pairs of images (Pix2Pix \cite{Isola2017} and CycleGAN \cite{Zhu2017}) as well as content/style disentanglement methods such as Multimodal Unsupervised Image-to-Image Translation (MUNIT) \cite{munit}. Pix2Pix was able to generate high quality images using a conditional GAN but required paired and registered images to train effectively. Instead, CycleGAN can be trained using unpaired data by using two generators to map between each domain. 
%Similarly to Pix2Pix, discriminators are used to classify between real and fake images. 
% A cycle consistency loss is also used to ensure the learning of mappings that preserve the relevant content of the input image, whilst transforming it into the target domain. 
However, both Pix2Pix and CycleGAN struggle to generate diverse outputs when new style images are introduced at inference due to learning to map between domains directly. 

MUNIT offers an alternative approach using content/style disentanglement to take input images and predict latent spaces for style and content encoding. This allows style transfer whilst preserving the content of the image, giving better control for the translation process. However, MUNIT often requires a large amount of data to train as well as having large computational and memory requirements. AdaIN is also used to combine the style and content latent spaces. It can, however, result in global rather than tissue dependent enhancement which leads to  the generation of unrealistic images. Additionally, MUNIT does not utilise any temporal information available in the DCE-MRI sequence. 

DRIT++ \cite{dritplusplus}, an extension to MUNIT, enhances diversity by introducing disentanglement at the domain level, and improving attribute manipulation. However, it also comes with increased computational complexity, complexity in hyperparameter tuning, and a potential trade-off between diversity and fidelity. 
% When generating DCE-MR images, we need to be able to produce images that are diverse as well as high fidelity, as we need the images to be realistic. 

Early work on style transfer for videos naively processed frames independently leading to flickering and false discontinuities \cite{chendong2017}.
% Style transfer has also been extended to videos, with early work naively processing frames independently. 
Methods have since been proposed to ensure temporal consistency between image frames \cite{chendong2017}. They use three components: a style sub-network, a flow sub-network and a mask sub-network. The style sub-network is an auto-encoder which performs the style transfer. The flow sub-network estimates the correspondence between consecutive image frames and warps the features. Finally, the mask sub-network regresses a mask to compose features adjacent in time, allowing for warped features to be reused. This allows for a smooth style transfer between image frames, but it is prone to errors when there is large motion between frames as well as propagating errors over time leading to inconsistent style transfer and blurriness. 

\section{Materials and Methods}
\subsection{Data}
Two datasets were used. The first contains 2D kidney DCE-MRI \cite{Lietzmann2012} from 13 patients with 375 images acquired continuously at a temporal resolution of 1.6 s, a spatial resolution of 384x348 with pixel sizes 1.08x1.08 mm. The second contains 3D prostate DCE-MRI \cite{lemaitre} from 20 patients with 40 volumes acquired continuously at a temporal resolution of 6 s, a spatial resolution of 256x192x16 with voxel sizes 1.12x1.12x3.5 mm. Each dataset was split at a patient level with an 80:20 ratio, training was done using five-fold cross validation.

\subsection{Global Architecture}
\begin{figure*}[!htb]
\centering
\begin{minipage}[b]{1.0\linewidth}
  \centering
  \centerline{\includegraphics[width=12cm]{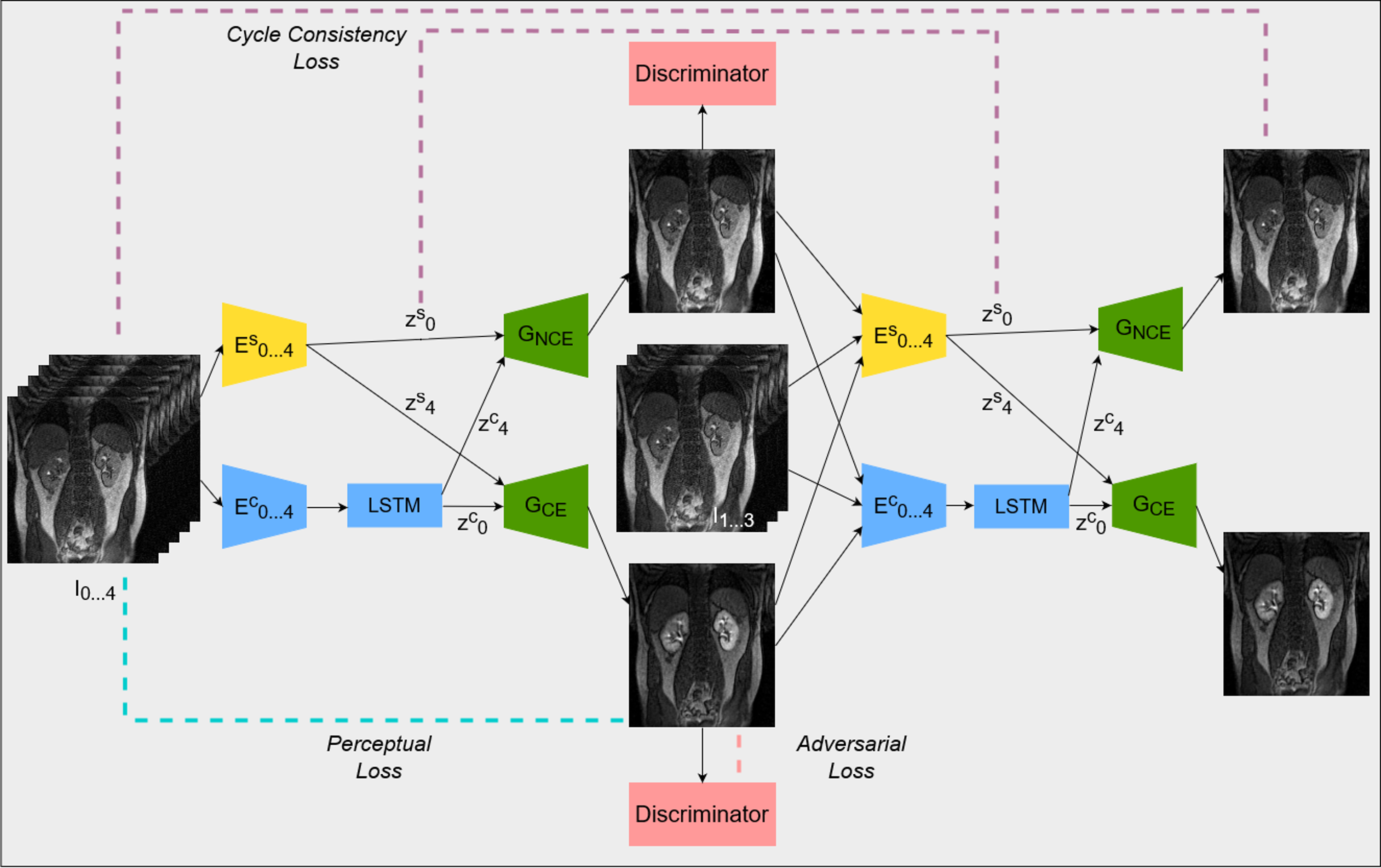}}
%  \vspace{2.0cm}
\end{minipage}
\caption{Our model takes an input \textit{I} of 5 images/volumes to content encoders E$^c_{0...4}$ and style encoders E$^s_{0...4}$. Generators G$_{CE}$ and G$_{NCE}$ construct images from latent spaces \textit{z} predicted by the content and style encoders. We also show the losses used: cycle consistency, perceptual and adversarial. For clarity of the figure, we have only shown one example for each of the losses.}
\label{fig:method}
\end{figure*}

Inspired by MUNIT, we use a structure composed of encoders and decoders to first disentangle content and style latent spaces and recombine them to generate new images. Successful disentanglement should lead to content latent spaces containing information representing the structures of the image and style latent spaces containing information representing the modality (MRI) and contrast enhancements. However, temporal information is not taken into account in most style transfer methods. Within DCE-MRI, information such as structures within organs become visible when there are various levels of contrast enhancement at different time points. In our method, we use additional content encoders to predict latent spaces for images from multiple time points. Then, we pass the predicted content latent spaces to a convolutional LSTM \cite{Chao2018} which has been set up to be bi-directional. We use a convolutional LSTM as this allows modelling of the temporal information from a sequence of images. It leverages spatial invariance by extracting relevant features from the images regardless of their position. Using a bi-directional LSTM allows better contextual understanding of the time series data by accessing past and future events. This allows content visible at various time points to be learnt. The overall architecture is shown in figure \ref{fig:method}. At inference the images are passed through the encoders and decoders once to perform style transfer.

\subsection{Instance Normalisation}

AdaIN \cite{Huang2017} is used to normalise the activations in a neural network based on the statistics of the style image. This enables the transfer of style from the style image to the content image by matching their statistical properties. However, by only using global style information, contrast enhancement is performed on the whole image instead of specific tissues. To overcome this, we use AdaConv \cite{Chandran2021} which captures global and local information to predict parameters to combine style and content latent spaces. Three convolutional neural networks are used to predict a depth-wise spatial kernel, a point-wise kernel and a bias. The kernels and bias are then convolved with the content code.

\subsection{Losses}
The L1 loss between the original and reconstructed images and between the initially predicted and reconstructed latent spaces is used to ensure that important information such as structure is not lost during the reconstruction and translation process. Furthermore, the L1 loss is known for its smooth gradients which can facilitate stable and effective optimisation during the training process. A mean squared error (MSE) loss is also used as an adversarial loss to help generate images that are indistinguishable from the real images.

While those two losses were used in MUNIT, we found that some important textural characteristics of the images were lost in the generated images. To improve this, we also used two additional perceptual loss functions proposed by Johnson \textit{et al.} \cite{justin16} for style transfer and super resolution tasks. The first is a feature reconstruction loss which is the squared, normalised Euclidean distance (Eq. \ref{eqn:percfeat1}). This encourages the feature representations predicted by a pretrained model \textit{P} of the content image \textit{x} and the generated image \textit{g} to be similar. \textit{f} is the dimension of the feature map.  

\begin{equation}
\label{eqn:percfeat1}
L_{feature} = \frac{1}{f}||P(g) - P(x)||^2_2
\end{equation}

Next, we use a style reconstruction loss to penalise differences in style between the style image \textit{y} and the generated image \textit{g}. This loss uses the squared Frobenius norm of the difference between the Gram matrices of the feature representations predicted by a pretrained network \textit{P} (Eq. \ref{eqn:percstyl1}). Minimising this loss encourages the generated image to capture style patterns and textures that are similar to the style image. The Frobenius norm is a measure of the magnitude of the matrix, and using it as the loss term allows the model to focus on preserving the overall style content and structure.

\begin{equation}
\label{eqn:percstyl1}
L_{Frob} = ||Gram(P(g)) - Gram(P(y))||^2_F
\end{equation}

For the 2D data, we used a VGG-16 pre-trained network on ImageNet \cite{ILSVRC15} converted to grayscale. For the 3D data we used MedicalNet which is a pretrained multi-modal multi-tissue ResNet-18 backbone.

\subsection{Evaluating style transfer}
Quantitatively evaluating style transfer is difficult with unpaired, unregistered data. Metrics such as peak signal to noise ratio (PSNR), SSIM and multiscale (MS)-SSIM primarily focus on the pixel-level similarity between a pair of images. This makes it difficult to evaluate style transfer in DCE-MRI due to the large variety of CE in an image and the localised aspect of it. We propose a contrast weighted (CW)-SSIM (Eq. \ref{eqn:wssim}) as a metric to measure both the content structure as well as evaluate the amount of CE (addition or removal) in the style image. 

For evaluating content, we weight the SSIM with a distance map \textit{dist\_map} created by calculating the shortest distance from each voxel to a contrast enhanced voxel. The distances were normalised between 0.1 and 1. We determined which voxels were contrast enhanced by subtracting each image from the first non-CE image and calculated the average. Then, we used an empirical threshold of $20$ to determine which voxels were contrast enhanced. 
% The distance map is either 2D or 3D depending on the number of dimensions of the data. 
To evaluate style, we inverted the distance map so that a voxel has a higher weighting when it is closer to a contrast enhanced voxel. \textit{x} is the generated image and \textit{y} is either the content or style image.

\begin{equation}
\label{eqn:wssim}
\begin{aligned}
% &\hat{x} = x \cdot \mathrm{dist\_map} , \quad \hat{y} = y \cdot \mathrm{dist\_map} \\
\mathrm{CW{\text -}SSIM}(\mathrm{x}, \mathrm{y}, \mathrm{dist\_map}) = \mathrm{SSIM(x \cdot \mathrm{dist\_map}, y \cdot \mathrm{dist\_map})}
\end{aligned}
\end{equation}

\subsection{Implementation}
%We used PyTorch to implement our approach and trained with the Adam optimiser,
We used PyTorch and trained with the Adam optimiser, a learning rate of 0.001 and batch size of 8 for 2D and 5 for 3D. Early stopping was used with a patience of 20. Our experiments were run on an RTX Titan GPU.
We compared our method with MUNIT, CycleGAN and StyleGAN3 \cite{stylegan3}. DRIT++ was not used due to the large computational requirements with 3D data.

\section{Results}
\begin{figure}[b!]
    \centering
    \hfill
    \begin{subfigure}[t]{0.23\textwidth}
        \includegraphics[width=\linewidth]{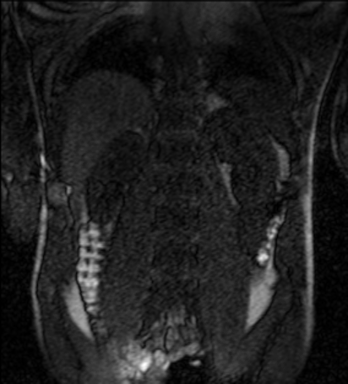}
        \caption{No CE Image}
        \label{fig:Kidney NoCE}
    \end{subfigure}
    \hspace{1mm}
    \begin{subfigure}[t]{0.23\textwidth}
        \includegraphics[width=\linewidth]{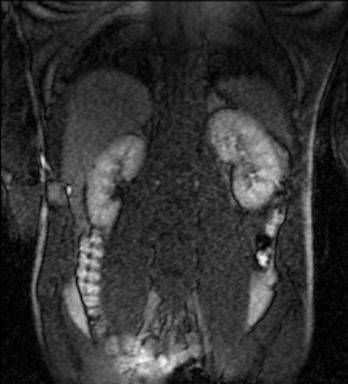}
        \caption{CE Image}
        \label{fig:Kidney CE}
    \end{subfigure}
    \hfill
    \vspace{1em}
    
    \begin{subfigure}[t]{0.23\textwidth}
        \includegraphics[width=\linewidth]{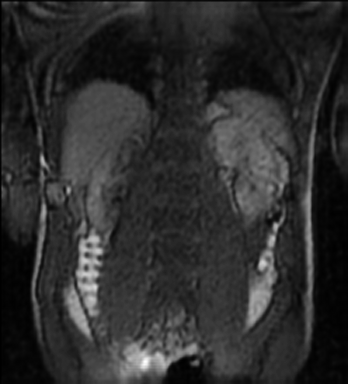}
        \captionsetup{font=scriptsize}
        \caption*{SSIM: 0.59 \\ (C) CW-SSIM: 0.71 \\ (S) CW-SSIM: 0.63}
        \label{fig:munitabkidney}
         
    \end{subfigure}
    \hfill
    \begin{subfigure}[t]{0.23\textwidth}
        \includegraphics[width=\linewidth]{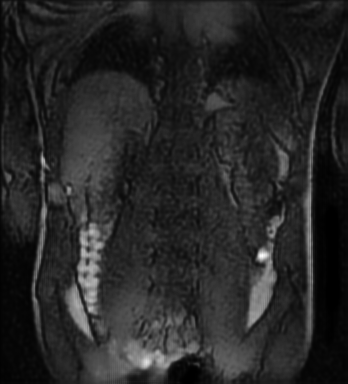}
        \captionsetup{font=scriptsize}
        \caption*{SSIM: 0.65 \\ (C) CW-SSIM: 0.59 \\ (S) CW-SSIM: 0.41}
        \label{fig:cycleabkidney}
    \end{subfigure}
    \hfill
    \begin{subfigure}[t]{0.23\textwidth}
        \includegraphics[width=\linewidth]{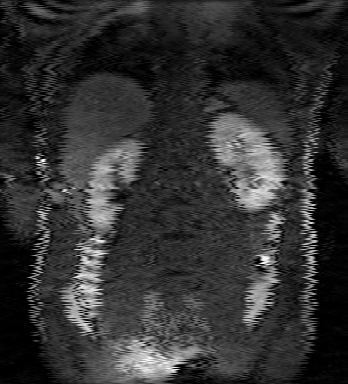}
        \captionsetup{font=scriptsize}
        \caption*{SSIM: 0.32 \\ (C) CW-SSIM: 0.44 \\ (S) CW-SSIM: 0.69}
        \label{fig:styleganabkidney}
    \end{subfigure}
    \hfill
    \begin{subfigure}[t]{0.23\textwidth}
        \includegraphics[width=\linewidth]{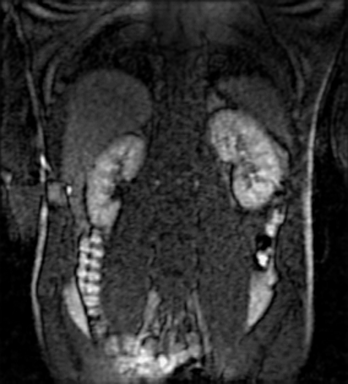}
        \captionsetup{font=scriptsize}
        \caption*{SSIM: 0.91 \\ (C) CW-SSIM: 0.94 \\ (S) CW-SSIM: 0.75}
        \label{fig:oursabkidney}
    \end{subfigure}
    
     \vspace{1em}
    
    \begin{subfigure}[t]{0.23\textwidth}
        \includegraphics[width=\linewidth]{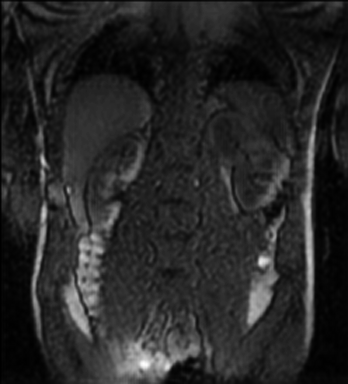}
        \captionsetup{font=scriptsize}
        \caption*{SSIM: 0.57 \\ (C) CW-SSIM: 0.68 \\ (S) CW-SSIM: 0.61}
        \captionsetup{font=footnotesize}
        \caption{MUNIT}
        \label{fig:munitbakidney}
    \end{subfigure}
    \hfill
    \begin{subfigure}[t]{0.23\textwidth}
        \includegraphics[width=\linewidth]{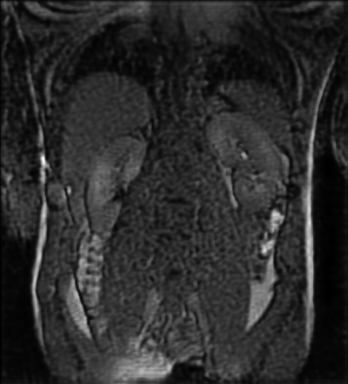}
        \captionsetup{font=scriptsize}
        \caption*{SSIM: 0.56 \\ (C) CW-SSIM: 0.54 \\ (S) CW-SSIM: 0.51}
        \captionsetup{font=footnotesize}
        \caption{CycleGAN}
        \label{fig:cyclebakidney}
    \end{subfigure}
    \hfill
    \begin{subfigure}[t]{0.23\textwidth}
        \includegraphics[width=\linewidth]{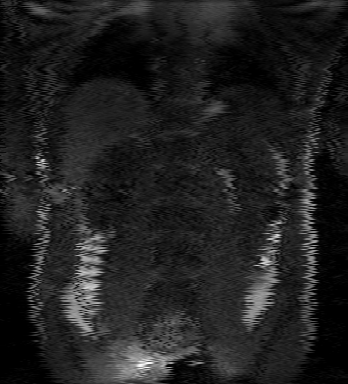}
        \captionsetup{font=scriptsize}
        \caption*{SSIM: 0.31 \\ (C) CW-SSIM: 0.42 \\ (S) CW-SSIM: 0.67}
        \captionsetup{font=footnotesize}
        \caption{StyleGAN3}
        \label{fig:styleganbakidney}
    \end{subfigure}
    \hfill
    \begin{subfigure}[t]{0.23\textwidth}
        \includegraphics[width=\linewidth]{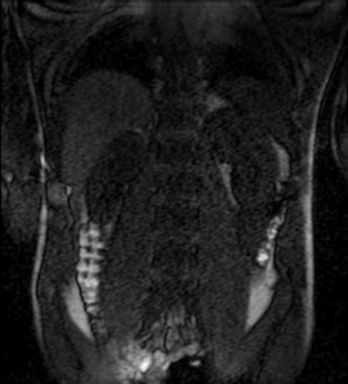}
        \captionsetup{font=scriptsize}
        \caption*{SSIM: 0.89 \\ (C) CW-SSIM: 0.93 \\ (S) CW-SSIM: 0.73}
        \captionsetup{font=footnotesize}
        \caption{Our Method}
        \label{fig:oursbakidney}
    \end{subfigure}
    
    \caption{Example results from different style transfer approaches with the kidney data. The first row (a and b) shows the input images. The second row shows the results when (a) is used as the content image and (b) as the style. The third row shows results when (b) is used as the content image and (a) as the style. We also show scores given by the SSIM, content (C) CW-SSIM and style (S) CW-SSIM.}
    \label{fig:resultskidney}
\end{figure}

\begin{figure}
    \centering
    \hfill
    \begin{subfigure}[t]{0.24\textwidth}
        \includegraphics[width=\linewidth]{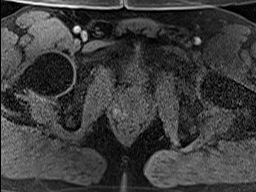}
        \caption{No CE Image}
        \label{fig:prostate NoCE}
    \end{subfigure}
    \begin{subfigure}[t]{0.24\textwidth}
        \includegraphics[width=\linewidth]{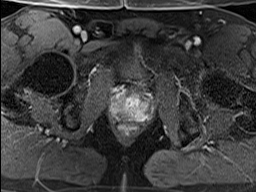}
        \caption{CE Image}
        \label{fig:prostate CE}
    \end{subfigure}
    \hfill
    \vspace{1em}
    
    \begin{subfigure}[t]{0.24\textwidth}
        \includegraphics[width=\linewidth]{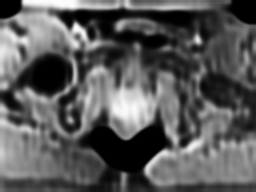}
        \captionsetup{font=scriptsize}
        \caption*{SSIM: 0.41 \\ (C) CW-SSIM: 0.39 \\ (S) CW-SSIM: 0.29}
        \label{fig:munitabprostate}
    \end{subfigure}
    \hfill
    \begin{subfigure}[t]{0.24\textwidth}
        \includegraphics[width=\linewidth]{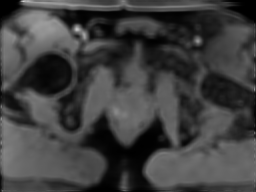}
        \captionsetup{font=scriptsize}
        \caption*{SSIM: 0.44 \\(C) CW-SSIM: 0.53 \\ (S) CW-SSIM: 0.31}
        \label{fig:cycleabprostate}
    \end{subfigure}
    \hfill
    \begin{subfigure}[t]{0.24\textwidth}
        \includegraphics[width=\linewidth]{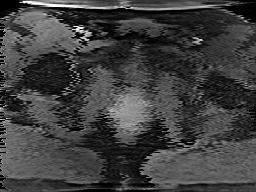}
        \captionsetup{font=scriptsize}
        \caption*{SSIM: 0.32 \\ (C) CW-SSIM: 0.42 \\ (S) CW-SSIM: 0.54}
        \label{fig:styleganabprostate}
    \end{subfigure}
    \hfill
    \begin{subfigure}[t]{0.24\textwidth}
        \includegraphics[width=\linewidth]{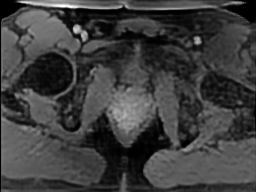}
        \captionsetup{font=scriptsize}
        \caption*{SSIM: 0.69 \\ (C) CW-SSIM: 0.93 \\ (S) CW-SSIM: 0.62}
        \label{fig:oursabprostate}
    \end{subfigure}
    
    \vspace{1em}
    
    \begin{subfigure}[t]{0.24\textwidth}
        \includegraphics[width=\linewidth]{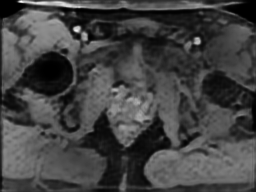}
        \captionsetup{font=scriptsize}
        \caption*{SSIM: 0.71 \\ (C) CW-SSIM: 0.61 \\ (S) CW-SSIM: 0.52}
        \captionsetup{font=footnotesize}
        \caption{MUNIT}
        \label{fig:munitbaprostate}
    \end{subfigure}
    \hfill
    \begin{subfigure}[t]{0.24\textwidth}
        \includegraphics[width=\linewidth]{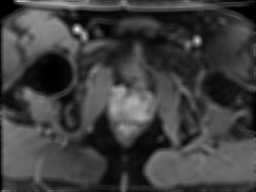}
        \captionsetup{font=scriptsize}
        \caption*{SSIM: 0.61 \\ (C) CW-SSIM: 0.42 \\ (S) CW-SSIM: 0.19}
        \captionsetup{font=footnotesize}
        \caption{CycleGAN}
        \label{fig:cyclebaprostate}
    \end{subfigure}
    \hfill
    \begin{subfigure}[t]{0.24\textwidth}
        \includegraphics[width=\linewidth]{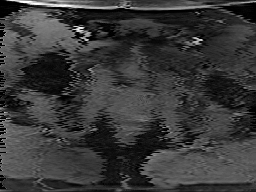}
        \captionsetup{font=scriptsize}
        \caption*{SSIM: 0.42 \\ (C) CW-SSIM: 0.28 \\ (S) CW-SSIM: 0.49}
        \captionsetup{font=footnotesize}
        \caption{StyleGAN3}
        \label{fig:styleganbaprostate}
    \end{subfigure}
    \hfill
    \begin{subfigure}[t]{0.24\textwidth}
        \includegraphics[width=\linewidth]{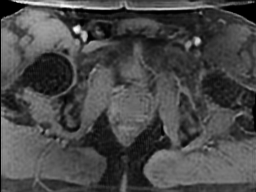}
        \captionsetup{font=scriptsize}
        \caption*{SSIM: 0.79 \\ (C) CW-SSIM: 0.92 \\ (S) CW-SSIM: 0.59}
        \captionsetup{font=footnotesize}
        \caption{Our Method}
        \label{fig:oursbaprostate}
    \end{subfigure}
    
    \caption{Example results from different style transfer approaches with the prostate data. The first row (a and b) shows the input images. The second row shows the results when (a) is used as the content image and (b) as the style. The third row shows results when (b) is used as the content image and (a) as the style. We also show scores given by the SSIM, content (C) CW-SSIM and style (S) CW-SSIM.}
    \label{fig:resultsprostate}
\end{figure}

Tables \ref{tab:kidney} and \ref{tab:prostate} show our quantitative results: PSNR between the style (image we want to transfer style from) and generated image, SSIM and MS-SSIM between the content (image we want to take structure from) and generated image and finally, our proposed weighted SSIMs. For each style transfer direction and metric, our method consistently outperforms the other approaches. Figures \ref{fig:resultskidney} and \ref{fig:resultsprostate} qualitatively highlight the good results of our method on 2D and 3D datasets. In addition, quantitative results on both figures show that our proposed metric preserves qualitative ordering of results for both style and content. In comparison, metrics such as SSIM struggle to accurately correlate to visual results.

% For adding CE in the kidney, we rank in descending order for content quality: our method, MUNIT, CycleGAN and StyleGAN3. The SSIM score places CycleGAN before MUNIT which does not follow the qualitative results, whereas our content CW-SSIM does follow the qualitative results. For style quality, we rank in descending order: our method, StyleGAN3, MUNIT and CycleGAN. The SSIM scores again do not follow this order, whereas the style CW-SSIM does follow the qualitative results. For the other directions, 

\begin{table}[b!]
\caption{Quantitative results for each approach when adding or removing CE in the kidney. The mean and standard deviation is shown for each metric.} 
  \centering
  \begin{tabular}{|c|c|c|c|c|c|}
    \hline
    \textbf{Metric} & \textbf{Direction} & \textbf{MUNIT} & \textbf{CycleGAN} & \textbf{StyleGAN3} & \textbf{Our Method} \\
    \hline
    \multirow{2}{*}{\textbf{PSNR}} & Non-CE to CE & 70.6 $\pm$ 3.2 & 66.5 $\pm$ 2.7 & 52.3 $\pm$ 3.9 & \textbf{ 81.3 $\pm$ 2.7} \\
    \cline{2-6}
     & CE to Non-CE & 71.5 $\pm$ 3.4 & 66.8 $\pm$ 2.8 & 51.9 $\pm$ 3.8 & \textbf{ 78.5 $\pm$ 2.6} \\
    \hline

    \multirow{2}{*}{\textbf{SSIM}} & Non-CE to CE & 0.61 $\pm$ 0.05 & 0.66 $\pm$ 0.06 & 0.32 $\pm$ 0.06 & \textbf{0.89 $\pm$ 0.04} \\
    \cline{2-6}
     & CE to Non-CE & 0.58 $\pm$ 0.04 & 0.58 $\pm$ 0.05 & 0.34 $\pm$ 0.07 &\textbf{0.88 $\pm$ 0.03}  \\
    \hline
    
    \multirow{2}{*}{\textbf{MS-SSIM}} & Non-CE to CE & 0.55 $\pm$ 0.04 & 0.43 $\pm$ 0.07 & 0.34 $\pm$ 0.04 & \textbf{ 0.82 $\pm$ 0.03 } \\
    \cline{2-6}
     & CE to Non-CE & 0.52 $\pm$ 0.04 & 0.45 $\pm$ 0.05 & 0.37 $\pm$ 0.05 & \textbf{ 0.79 $\pm$ 0.04 } \\
    \hline
    
    \textbf{Content} & Non-CE to CE & 0.73 $\pm$ 0.03 & 0.61 $\pm$ 0.05 & 0.47 $\pm$ 0.05 & \textbf{ 0.95 $\pm$ 0.02 } \\
    \cline{2-6}
    \textbf{CW-SSIM} & CE to Non-CE & 0.69 $\pm$ 0.03 & 0.57 $\pm$ 0.05 & 0.41 $\pm$ 0.05 & \textbf{ 0.94 $\pm$ 0.02 } \\
    \hline
    \textbf{Style} & Non-CE to CE & 0.64 $\pm$ 0.03 & 0.41 $\pm$ 0.05 & 0.67 $\pm$ 0.05 & \textbf{ 0.74 $\pm$ 0.02 } \\
    \cline{2-6}
    \textbf{CW-SSIM} & CE to Non-CE & 0.62 $\pm$ 0.03 & 0.49 $\pm$ 0.05 & 0.64 $\pm$ 0.05 & \textbf{ 0.72 $\pm$ 0.02 } \\
    \hline
  \end{tabular}
  \label{tab:kidney}
\end{table}

%%%%%%%%%%%%%%%%%%%%%%%%%%
% Kidney results
%%%%%%%%%%%%%%%%%%%%%%%%%%

%%%%%%%%%%%%%%%%%%%%%%%%%%
% Prostate results
%%%%%%%%%%%%%%%%%%%%%%%%%%

\begin{table}
\caption{Quantitative results for each approach when adding or removing CE in the prostate. The mean and standard deviation is shown for each metric.} 
  \centering
  \begin{tabular}{|c|c|c|c|c|c|}
    \hline
    \textbf{Metric} & \textbf{Direction} & \textbf{MUNIT} & \textbf{CycleGAN} & \textbf{StyleGAN3} & \textbf{Our Method} \\
    \hline
    \multirow{2}{*}{\textbf{PSNR}} & Non-CE to CE & 40.6 $\pm$ 4.3 & 52.2 $\pm$ 3.8 & 43.5 $\pm$ 5.2 & \textbf{72.1 $\pm$ 3.6} \\
    \cline{2-6}
     & CE to Non-CE & 68.7 $\pm$ 4.1 & 61.2 $\pm$ 3.4 & 51.9 $\pm$ 6.1 & \textbf{68.9 $\pm$ 4.1} \\
    \hline
    \multirow{2}{*}{\textbf{SSIM}} & Non-CE to CE & 0.41 $\pm$ 0.06 & 0.46 $\pm$ 0.05 & 0.36 $\pm$ 0.09 & \textbf{0.71 $\pm$ 0.04} \\
    \cline{2-6}
     & CE to Non-CE & 0.69 $\pm$ 0.04 & 0.59 $\pm$ 0.05& 0.43 $\pm$ 0.08 &\textbf{0.78 $\pm$ 0.03}\\
    \hline
    \multirow{2}{*}{\textbf{MS-SSIM}} & Non-CE to CE & 0.49 $\pm$ 0.04 & 0.38 $\pm$ 0.06 & 0.28 $\pm$ 0.05 & \textbf{ 0.76 $\pm$ 0.04 } \\
    \cline{2-6}
     & CE to Non-CE & 0.47 $\pm$ 0.04 & 0.39 $\pm$ 0.05 & 0.32 $\pm$ 0.06 & \textbf{ 0.71 $\pm$ 0.03 } \\
    \hline
    \textbf{Content} & Non-CE to CE & 0.38 $\pm$ 0.04 & 0.52 $\pm$ 0.05 & 0.41 $\pm$ 0.05 & \textbf{0.92 $\pm$ 0.03} \\
    \cline{2-6}
    \textbf{CW-SSIM} & CE to Non-CE & 0.63 $\pm$ 0.04 & 0.46 $\pm$ 0.04 & 0.32 $\pm$ 0.06 & \textbf{0.93 $\pm$ 0.04} \\
    \hline
    \textbf{Style} & Non-CE to CE & 0.34 $\pm$ 0.05 & 0.28 $\pm$ 0.06 & 0.52 $\pm$ 0.07 & \textbf{0.59 $\pm$ 0.05} \\
    \cline{2-6}
    \textbf{CW-SSIM} & CE to Non-CE & 0.53 $\pm$ 0.04 & 0.21 $\pm$ 0.05 & 0.47 $\pm$ 0.06 & \textbf{0.57 $\pm$ 0.04} \\
    \hline
  \end{tabular}
  \label{tab:prostate}
\end{table}

\section{Discussion}
In this work, we proposed a method for style transfer that leverages spatio-temporal information, applied to DCE-MRI. Using convolutional LSTMs we are able to model changes in structures caused by CE across the temporal dimension giving better content latent spaces. In addition, using localised style information instead of a global one by making use of AdaConv helped to tackle the local nature of CE in DCE-MRI. 

Qualitative evaluation shows that the proposed method leads to sharper images, better content preservation, better localised CE and realistic MRI appearance. Quantitatively, we outperform the other algorithms with each metric, for each style transfer direction. While we expected that adding CE to images would be an easier task than removing CE, we found that for both tasks the method exhibits similarly performing quantitative and qualitative results. The results for style transfer with the kidney data show that when there is clear CE, it is easier to perform style transfer in both directions. However, when style transfer is performed on the prostate data, the model struggles in comparison to the kidney data. This may be due to the enhancement of the prostate being less defined than the kidney data. In figure \ref{fig:prostate CE}, we can see the prostate with CE. Compared to figure \ref{fig:Kidney CE} showing kidneys with CE, it is harder to determine the edges. The two original images shown in figures \ref{fig:prostate NoCE} and \ref{fig:prostate CE} are clear and easy to see some edges of prostate, in other images, it is harder to see the edges of the prostate. 

Our metric, CW-SSIM, enables the evaluation of the quality of the content and style of the generated images when there are large differences in CE. When comparing against the qualitative results, the quality of the style transfer is in line with each of the CW-SSIM results. By using two weightings: areas of CE and without CE, we are able to evaluate the similarity between structures where intensity differences are small to measure for content similarity. By using an inverted map, we can also compare regions of high CE to measure the differences in style between the generated and style image. An advantage of this metric over SSIM or MS-SSIM is that we are able to separate the evaluation of content and style to ensure that key structures in the image remain, whilst changing the intensity values to match the style. This is important with DCE-MRI as large intensity changes are localised in specific tissues. When we compare the results for the prostate using MS-SSIM and style CW-SSIM, we can see that the style CW-SSIM gave a much lower score. This follows the qualitative results closely as the styles of each of the generated images are not similar to the style image used. By evaluating the areas of the image separately we remove the possibility of parts of the image with little CE inflating the score given by the SSIM.

A limitation to our style transfer approach is that better style transfer was achieved with the kidney data compared to the prostate data. This may be due to the structures in the prostate data not being as clearly defined as the kidney data, as well as the prostate not enhancing as clearly as the kidney. Similarly to MUNIT, our method has large computational and memory requirements.
Using our method, new images can be generated, to contain a variety of stages of CE. This type of augmentation allows a small number of images to be manually annotated making it easier to train robust models for image processing tasks.

\section{Conclusion}
We proposed a style transfer method which used temporal information with disentangled representations. Our method learns information from time series data and accurately captures key structures from content images whilst localising the addition of CE. Our method is trained in an unsupervised setting with unpaired data. Qualitative and quantitative analysis shows that our method outperforms other popular style transfer techniques. We also proposed a new metric which uses information based on the localised aspect of tissue enhancement with contrast in DCE-MRI to evaluate similarity to the style or content image.

\section{Acknowledgements}
This work was funded by Medical Research Scotland and CMRE.

\bibliographystyle{splncs04}

\bibliography{refs}

\end{document}